\documentclass[preprint, journal]{vgtc}             

\usepackage{enumitem}
\usepackage{amsmath}
\usepackage[pagebackref,bookmarks]{hyperref}  
\usepackage{multirow}
\usepackage{graphicx}
\usepackage{xspace}

\newcommand{\ND}{\textbf{\texttt{ND}}\xspace}
\newcommand{\CD}{\textbf{\texttt{CD}}\xspace}
\newcommand{\ID}{\textbf{\texttt{ID}}\xspace}

\preprinttext{To Appear in 2025 TVCG Special Issue on the 2025 IEEE Conference on Virtual Reality and 3D User Interfaces (VR)}

\vgtccategory{Research}
\vgtcpapertype{theory/model}

\title{Reaction Time as a Proxy for Presence in Mixed Reality with Distraction}

\author{Yasra Chandio\thanks{e-mail: ychandio@umass.edu}\\ %
        \scriptsize University of Massachusetts Amherst\\
\and  Victoria Interrante\thanks{e-mail: interran@umn.edu}\\ %
     \scriptsize University of Minnesota Twin Cities %
\and Fatima M. Anwar \thanks{e-mail: fanwar@umass.edu}\\ %
     \scriptsize University of Massachusetts Amherst %
     }

\abstract{%
Distractions in mixed reality (MR) environments can significantly influence user experience, affecting key factors such as presence, reaction time, cognitive load, and Break in Presence (BIP). Presence measures immersion, reaction time captures user responsiveness, cognitive load reflects mental effort, and BIP represents moments when attention shifts from the virtual to the real world, breaking immersion. While prior work has established that distractions impact these factors individually, the relationship between these constructs remains underexplored, particularly in MR environments where users engage with both real and virtual stimuli.
To address this gap, we have presented a theoretical model to understand how congruent and incongruent distractions affect all these constructs. We conducted a within-subject study ($N=54$) where participants performed image-sorting tasks under different distraction conditions. Our findings show that incongruent distractions significantly increase cognitive load, slow reaction times, and elevate BIP frequency, with presence mediating these effects.

}


\teaser{
    \centering
    \vspace{-0.4cm}
    \begin{tabular}{c}
    \includegraphics[width=\textwidth]{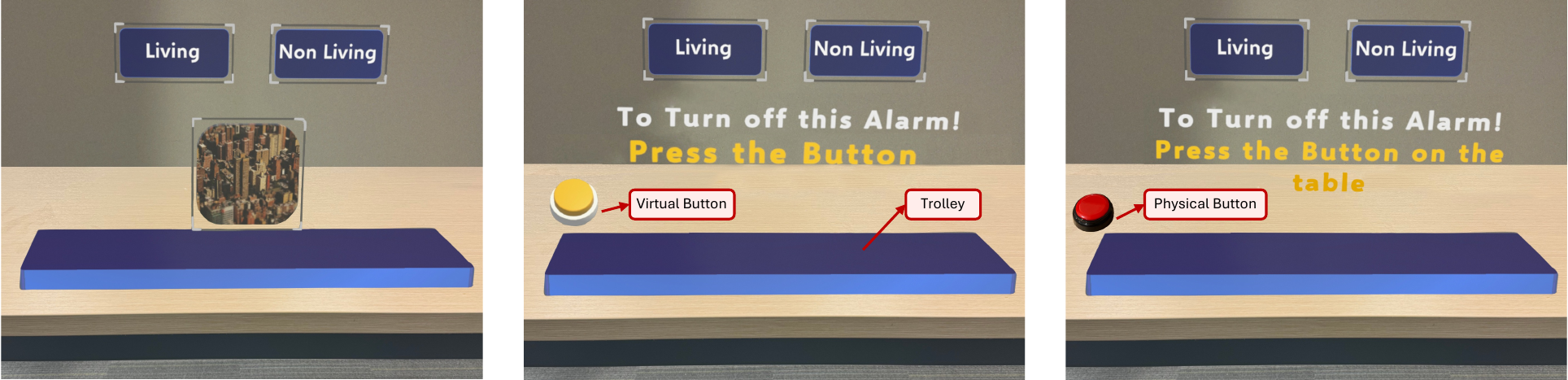} 
    \end{tabular}
    \caption{\emph{View of the virtual scene across three MR conditions: In No Distraction (\ND), participants sort images as living or non-living without distractions. In Congruent Distraction (\CD), they sort images with virtual distractions. In Incongruent Distraction (\ID), they sort images with real-world distractions. Red boxes and arrows label scene elements and are not part of the experiment.}}
  \label{fig:teaser}
}

\usepackage{tabu}                      
\usepackage{booktabs}                  %
\usepackage{mwe}                       
\usepackage{flushend}
\usepackage{mathptmx}                  

\begin{document}

\firstsection{Introduction}
\maketitle
\label{sec:introduction}
In Mixed Reality (MR) environments, \emph{presence} refers to the immersive experience of being fully situated (\emph{``being there''}) in a virtual environment (VE)~\cite{presence-1, presence-2}. A high sense of presence is a critical factor for determining user immersion~\cite{gonzalez2017immersive}, engagement\cite{skarbez2020immersion}, task performance~\cite{presence-human-performance}, and interaction quality~\cite{bowman2001introduction-interaction-survey}. 
Presence is largely constructed through cognitive and perceptual processes~\cite{place-illusion-plausibility, plausible-tvcg-23}. Cognition shapes how users interpret and engage with the VE, while perception involves the sensory inputs that maintain the coherence of that experience. However, disruptions in either the real-world (noises, touch, etc.) or VE (system lags, glitches etc.) can affect these processes, often causing inconsistencies between the user's mental model and the VE~\cite{liebold2017continuous-bip-measure}, leading to the \textbf{\emph{break in presence (BIP)}}~\cite{steed2005breaks-usablity-bip, slater2000virtual-BIP-presence-counter}, causing the user's attention to shift from the VE back to real-world. Real-world elements inherently differ from virtual stimuli in their unpredictability and sensory modality, requiring users to reallocate cognitive attention across domains~\cite{liebold2017continuous-bip-measure}. This perceptual discontinuity may increase the likelihood of BIP and reduce presence.

Traditionally, subjective methods like post-experience questionnaires have been used to measure presence~\cite{presence-measure-subjective-objective}. These methods, however, have significant limitations. They fail to capture the real-time, moment-to-moment fluctuations in presence~\cite{in-vr-questionarrie} and are subject to biases in how participants recall or interpret their experiences~\cite{choquestionarrie2003dichotomy}. This makes it difficult to objectively measure presence as it dynamically changes throughout the MR experience. Prior studies have examined the correlation between presence and reaction time by manipulating perceptual factors such as object realism and plausibility~\cite{chandio-tvcg-23} and the link between the user's cognitive state and performance to assess presence through reaction time~\cite{chandio-vr-24-human-factors}. 

While the correlation between presence and reaction time has been established, less is known about how this relationship is affected by distractions, particularly in MR environments where users remain connected to the physical world. These environments introduce additional complexity because real or virtual distractions can disrupt a user’s attention~\cite{interruptions-presence-workload-attention, presence-behavior-attention-memory} and potentially lead to a BIP~\cite{break-in-presence}. In MR, distractions are not just an inevitable part of the experience; they also provide a key challenge for maintaining consistent presence~\cite{gamers-enhanced-cognitive-ablities-green2003action, chen2022predicting-attention-distractions} and low reaction times~\cite{reaction-time-cognition}. Distractions in MR environments can be categorized as congruent (aligned with the task) or incongruent (deviating from the task)~\cite{breaking-experience, plausible-tvcg-23}.
\emph{Congruent distractions} (\CD) might be more easily absorbed into the experience, but they still increase cognitive load as they require mental effort to integrate them into the ongoing task~\cite{cognitive-load-theory-sweller1988cognitive}. However, \emph{incongruent distractions} (\ID) are more likely to cause significant disruptions because they conflict with the user's task goals, forcing the brain to process unrelated information~\cite{wirth2007process-presence-distarction}. This mismatch between the distraction and the task increases cognitive load~\cite{odermatt2021congruency} and makes it more difficult to maintain focus, thereby increasing the likelihood of BIP~\cite{cognitiveload-presence}. These distractions can negatively impact users' reaction times~\cite{lok2003effects-interaction}, as their attention is forced to shift between real-world and virtual stimuli~\cite{wienrich2021spatial-presence-model-distractions}. To address this gap in the literature, we conducted an exploratory within-subjects study (N = 54). We evaluated the relationship between presence, cognitive load, BIP, and reaction time across \CD, \ID, and no-distraction (\ND) conditions caused by secondary tasks. We aim to explore how different types of distractions impact these key variables and how they interact in real-time MR environments.

 To explore this relationship further, we conducted an exploratory within-subjects study (N = 54). We evaluated the relationship between presence, cognitive load, BIP, and reaction time across congruent (\CD), incongruent (\ID), and no-distraction (\ND) conditions induced by secondary tasks.
Assuming that these secondary tasks successfully create distinct distraction scenarios, our study aims to address the following key research questions (RQ):
\begin{enumerate}[leftmargin=0.95cm, itemsep=-0.1cm, topsep=-0.1cm]
    \item[\textbf{RQ1:}] How do different types of distractions (congruent vs. incongruent) influence the likelihood of a BIP? 
    \item[\textbf{RQ2:}] What is the effect of distractions on cognitive load, and how does BIP interact with cognitive load? 
    \item[\textbf{RQ3:}] What is the relationship between presence and reaction time, and how does cognitive load mediate this relationship? 
\end{enumerate}

\section{Background and Related Work}
\label{sec:background}
\subsection{Terminologies}
In this paper, when we refer to \textbf{MR}, we align with the definition based on Milgram’s reality-virtuality continuum, positioning it near Augmented Reality (AR), where virtual components blend naturally with the physical world. 

\noindent \textbf{Presence} refers to the psychological sense of ``being there" in a VE, requiring visual realism and natural interactions~\cite{presence-1}.

\noindent \textbf{Cognitive Load} refers to the mental effort required to process information and perform tasks. 

\noindent \textbf{Reaction time} represents the time it takes for a user to respond to a specific stimulus in the VE, distinct from the total time required to complete an entire task, often referred to as task completion time; these terms are not used interchangeably for the rest of the paper.

\noindent \textbf{Immersion} refers to the objective quality of the system's ability to provide a coherent and rich sensory experience that fully engages the user~\cite{gonzalez2017immersive}. Immersion represents the degree of sensory fidelity, which influences but does not equate to presence. Cognitive and perceptual processes are the main factors influencing presence and our main focus in this paper.

\subsection{Measuring Presence}
\label{sec:measuring-presence}
Traditionally, presence has been measured using subjective self-reports. Popular questionnaires include the Presence Questionnaire (PQ)~\cite{ws}, Igroup presence questionnaire (IPQ)~\cite{ipq,ipq-2}, and others~\cite{itc-questionarrielessiter2001cross, IAS-questionarrie-leary1993interaction, BFI-10-questionarrie-rammstedt2007measuring, grassini2020questionnaire-review, breaking-experience, questionarie-prior-experience}. These tools evaluate the user's subjective sense of presence after engaging with the VE, asking participants to reflect on how immersed they felt. These questionnaires have been instrumental in understanding users' overall impressions of immersion, but they are limited by their retrospective nature, which can introduce bias and fail to capture real-time fluctuations in presence~\cite{in-vr-questionarrie}. 
The second challenge with subjective questionnaires in MR is that many AR/MR-specific questionnaires~\cite{gandy2010experiences-AR-questionnarie, georgiou2017development-AR-queststionnarie, hpq-ar-questionnaire, AR-presence-Regenbrecht2021MeasuringPI, arobject-questionaire-stevens2000sense} lack the validation and reliability testing as their VR counterparts~\cite{presence-object-stevens2002putting}. These newer tools often have not undergone rigorous validation processes like Cronbach’s alpha, Kaiser-Meyer-Olkin tests, or factor analyses~\cite{garcia2012cognitive-stats-cronbach-kaiser, tran2024survey-presence-MR-why-not-enough}. Therefore, researchers frequently rely on well-established VR questionnaires, which, while not perfect, offer more reliable alternatives than unvalidated MR-specific tools~\cite{presence-mr}. Prior studies have shown that VR-developed questionnaires can still be effective in AR/MR contexts, especially when all users experience similar environments, even if they are not fully immersive~\cite{questionarrie-in-realities-usoh2000using, presence-non-immersive}.

Alternatively, researchers have explored objective measures of presence that offer real-time feedback. Physiological measures such as heart rate and electrodermal activity~\cite{physiological-measure-meehan} and neurological signals~\cite{Neurophysiological-presence-measure} have been proposed as potential indicators of presence. However, these measures can be intrusive and may disrupt the immersive experience by adding physical discomfort.
Another objective measure is reaction time, a behavioral metric reflecting cognitive and motor response to stimuli. Reaction time is less invasive than physiological measures and offers a way to track presence continuously throughout the MR experience~\cite{reaction-time-cognition, reaction-time-correaltion}. Faster reaction times are often associated with higher levels of presence, while slower reaction times may signal BIP~\cite{garau2004temporal0slater-BIP-qualitative}. Reaction time has been widely used in various fields to assess attention~\cite{presence-behavior-attention-memory} and decision-making~\cite{presence-performance-measures} and is gaining attention in MR research for its potential to serve as a proxy for real-time presence~\cite{slater2000virtual-BIP-presence-counter, chandio-tvcg-23, reaction-time-szczurowski2017measuring, human-reaction-time-arif2022human}.

\subsection{Break in Presence (BIP)}
Slater et al.~\cite{slater2000virtual-BIP-presence-counter} defined BIP as moments when users shift focus from the virtual to the real-world, causing a noticeable drop in presence~\cite{wirth2007process-presence-distarction, break-in-presence}. BIP can be triggered by the following elements:

\begin{enumerate}[label=(\alph*), leftmargin=0.5cm]  
    \item \textbf{External distractions.} are real-world interruptions such as ambient noises, physical disturbances, or system notifications. External distractions force the user to divide their attention between the real-world and the VE, increasing cognitive load and making it harder to maintain presence~\cite{steed2005breaks-usablity-bip}.
    \item \textbf{Internal inconsistencies.} Glitches, lags, or unrealistic behaviors in the VE can create perceptual dissonance, forcing the user to question the validity of the VE and breaking their mental model~\cite{slater2006analysis-physiological-BIP}, leading to BIP~\cite{slater2000virtual-BIP-presence-counter,vona2024collabrating-AR-misalignment-element}.
    \item \textbf{Cognitive overload.} occurs when users are overwhelmed by task complexity or by processing real and virtual stimuli simultaneously~\cite{hofer2020role-plausiblity-presence-some-bip}, or discontinuation for secondary task~\cite{breaking-experience}. This leads to slower reaction times, reduced performance, and a higher likelihood of BIP~\cite{interruptions-presence-workload-attention,breves2023cognitive-load-presence-cybersickness}.
\end{enumerate}

BIP has also been explored as an objective measure of presence~\cite{averbukh2021measurement-bip-presence-counter}, and capturing real-time fluctuations. Instead of relying solely on post-experience questionnaires, researchers can track presence dynamically by observing when BIPs occur. While BIP-specific validated tools are limited, drops in presence scores have been used as indirect indicators of BIP~\cite{slater-usoh-steed1994depth}. For instance, researchers like Slater et al.\cite{slater2000virtual-BIP-presence-counter} asked participants to say ``here" when experiencing a BIP, leading to the development of a presence counter, which uses a Markov chain to track BIP frequency\cite{averbukh2021measurement-bip-presence-counter}. However, reliance on self-reports introduces biases and delays, as participants had to be trained to recognize and report BIP events, potentially influencing their behavior.
Subsequent studies linked physiological responses, such as heart rate and skin conductance, to correlate with BIP events~\cite{slater2003physiological-BIP, physiological-measure-meehan}. However, while physiological measures offer a more objective approach for continuously monitoring presence without requiring active user input, they can be intrusive and difficult to implement in naturalistic settings.
Despite these limitations, BIP remains valuable for presence research, especially when combined with measures like reaction time.

\subsection{Distractions Causing Disruptions in MR}
\label{sec:distraction-background}
Disruptions in MR arise from the dual interaction of the real and virtual worlds, which can complicate how users process sensory information and manage cognitive resources. Cognitive load and perceptual consistency are key factors in maintaining presence~\cite{cognition-perception-latoschik_2022}, and disruptions in either of these domains can lead to a BIP~\cite{break-in-presence}.
Cognitive load theory~\cite{cognitive-load-theory-sweller1988cognitive} divides cognitive load into intrinsic (task-related), extraneous (task-unrelated), and germane (learning-related) types. In MR environments, extraneous cognitive load is often increased by distractions from the real-world, such as ambient noise or physical movements, which divert attention away from the virtual task~\cite{kocur2020effects-external-perception-cognitive-performance}. Incongruent stimuli, unrelated to the virtual task, demand additional mental processing and increase cognitive load, which can overwhelm the user's ability to stay immersed in the VE~\cite{odermatt2021congruency}.

Perception in MR relies on coherent sensory inputs from both the real and virtual worlds. Perceptual disruption, such as mismatched visual, auditory, or haptic feedback, can create a dissonance that undermines the user's sense of presence. These disruptions require users to reallocate cognitive resources to resolve the conflict between real and virtual stimuli, increasing the likelihood of a BIP~\cite{liebold2017continuous-bip-measure}. Perceptual disruptions are particularly problematic in MR, where users are constantly required to navigate and reconcile inputs from both the real and virtual worlds~\cite{vona2024collabrating-AR-misalignment-element}.

\subsection{BIP, Presence, Reaction time} 
As users experience BIP, their cognitive resources are redirected, which may result in slower reaction times~\cite{cognitiveload-presence}. Bracken et al.~\cite{bracken2014revisiting-reaction-time-secondary-task} found that reaction times increased during BIP events, showing that secondary task reaction time effectively measures disruptions in presence and attention.  
George et al.~\cite{interruptions-presence-workload-attention} found that interruptions in VR increase cognitive workload, disrupting presence and impairing reaction time, a key concern in MR where both real and virtual interruptions affect performance~\cite{presence-performance-measures}. Schrader~\cite{schrader2012influence-presence-cognitive-load} expanded on this by examining how presence influences cognitive load and task performance, finding that increased cognitive load leads to slower reaction times and immersion. Breves and Stein~\cite{breves2023cognitive-load-presence-cybersickness} demonstrated how cognitive load can disrupt presence. Additionally, Nordahl et al.~\cite{nordahl2010distraction-as-measure-presence} proposed using controlled distractions as a measure of presence, showing that the ability of a system to keep users engaged despite distractions could be an effective indicator of immersion. Payne et al.~\cite{payne2024considerations-presence-secondary-tasks} provided a framework for understanding the impact of secondary tasks on immersion in VE, which we extend by focusing on MR environments and scope down to use it as a tool to cause distractions. 

\subsection{Contribution Beyond Related Work}
While previous studies have explored presence, BIP, cognitive load, and reaction time independently, none have examined how these factors intersect in MR environments. This study offers several key contributions: (1) We introduce the theoretical cognitive distraction model for MR, explaining how distractions in MR could impact users (\S\ref{sec:approach}). (2) We empirically demonstrate the distinct effects of distractions on cognitive load, reaction time, and presence (\S\ref{sec:study}). (3) We highlight reaction time as a real-time proxy for presence, offering a non-intrusive method to assess immersion (\S\ref{sec:eval}). Unlike prior studies that linked reaction time to presence without considering the impact of distractions, our study adds depth by exploring how distractions modulate the relationship between reaction time and presence (\S\ref{sec:disc}), particularly in environments with both virtual and real-world stimuli.

\vspace{-0.1cm}
\section{Approach}
\label{sec:approach}
\subsection{Conceptual Model}
\label{sec:main-model}
\begin{figure}[t]
    \centering
    \includegraphics[width=\columnwidth]{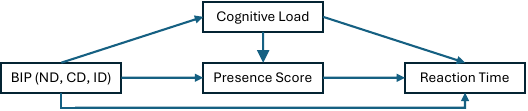}
    \vspace{-0.7cm}
    \caption{The Cognitive Distraction Model for MR (CDM-MR).}
    \label{fig:main-model}
\end{figure}

In this section, we present our theoretical \emph{Cognitive Distraction Model for MR (CDM-MR)}  and integrate the concepts discussed in the background to explain how distractions influence the user experience in MR, as shown in~\autoref{fig:main-model}.

\subsubsection{Cognitive Load and Attention}
To understand how distractions impact presence and BIP, we first examine the role of cognitive load and attention. In MR environments, managing both virtual and real-world stimuli requires significant cognitive effort, particularly when distractions are present. We assume that real-world stimuli are inherently more disruptive due to their sensory modality and external origin; however, individual differences in attention switching may vary.
Cognitive load theory~\cite{cognitive-load-theory-sweller1988cognitive} suggests that distractions increase extraneous cognitive load; because of that, when distraction occurs, users have fewer resources to maintain a presence in the VE~\cite{cognitiveload-presence}. This reflects the user’s need to allocate mental resources across multiple competing sources of information. This is supported by attention theory~\cite{ninio1974reaction-time-attention-theory}, which posits that human attentional capacity is finite. When users are faced with increasing cognitive demands,  particularly when it occurs abruptly due to unexpected distractions, their ability to maintain attention on a primary task may be reduced. This aligns with the concept of dual-task interference~\cite{pashler1994dual-task-interference}, which suggests that performing multiple tasks concurrently can lead to performance declines as attentional resources are divided between tasks, especially when the two competing tasks are conflicting, such as often incongruent nature of transitioning between real and virtual inputs, this task-switching incur cognitive costs due to attention reallocation~\cite{monsell2003task-switiching}. 

\subsubsection{Reaction Time and Presence}
BIP itself introduces additional cognitive load. Once a BIP occurs, users must expend mental effort to reorient themselves back into the VE, thereby increasing the cognitive load beyond what is required for the primary task alone. We hypothesize that this creates a feedback loop: distractions increase cognitive load, leading to BIP, further exacerbating cognitive load as the user attempts to regain immersion in the virtual world. As the cognitive resources are redirected due to BIP, it may result in slower reaction times~\cite{bracken2014revisiting-reaction-time-secondary-task}. This concept is further supported by Baumeister's resource depletion model~\cite{baumeister2000ego-resource-depletion}, which suggests that cognitive resources are finite, and each competing task or distraction (in our case, BIP) depletes the available attentional capacity for the primary task. In MR environments, frequent BIP events could lead to decreased presence and degraded performance due to the increased mental effort required to re-engage with the virtual world~\cite{cognitiveload-presence}.

\subsubsection{Distractions Types}
\vspace{-0.05cm}
In \S\ref{sec:distraction-background}, we discussed how distractions in MR affect users cognitively and perceptually; now we will discuss the nature of distractions by categorizing them into two distinct types: 

\vspace{0.1cm}
\noindent \textbf{Congruent distractions (\CD).} align with or are either contextually relevant to the task being performed in the VE or integrated within the VE to initiate parallel secondary tasks. 
For example, imagine a user in an MR environment navigating through a virtual space while following virtual cues. \CD might involve a virtual pop-up or instruction that fits naturally within the virtual task, such as an additional navigation prompt or task-related feedback. This type of distraction is integrated within the VE and aligns with the user’s mental model of the task, making it easier to absorb without breaking the illusion of immersion. The assumption is that the likelihood of a BIP is lower because the distraction is part of the VE.

\vspace{0.1cm}
\noindent \textbf{Incongruent distractions (\ID).} involve stimuli that break the illusion of immersion and repeatedly pull the user back to the real world. For example, an unrelated real-world notification, a sudden noise, or a secondary task that requires interactions in the real-world would force the user to shift their attention away from the virtual space. These distractions break immersion and increase the likelihood of BIP by diverting attention away from the virtual task and drawing focus back to the real world.
The critical difference between \CD and \ID lies in the cognitive and attentional demands they impose. In MR, \CD may add complexity but is easier to incorporate into the task structure, whereas \ID disrupts cognitive flow by introducing unrelated real-world interactions and creating this back-and-forth between virtual and real words, making it harder for users to maintain a presence.

\subsection{Hypotheses}
\vspace{-0.1cm}
Building on the \emph{CDM-MR}, our experimental design will focus on testing the following hypotheses:

\begin{itemize}[itemsep=-0.05cm, topsep=-0.1cm,leftmargin=0.35cm]
    \item \textbf{H1}: Incongruent distractions cause BIP, while congruent distractions do not cause BIP.
    \item \textbf{H2}: Distractions introduced through secondary tasks increase cognitive load.
      \begin{itemize}[itemsep=-0.05cm, topsep=-0.1cm,leftmargin=0.35cm]
       \item \textbf{H2a}: Incongruent distractions result in a greater increase in cognitive load compared to congruent distractions.
    \item \textbf{H2b}: BIP leads to an increase in cognitive load.
    \item \textbf{H2c}: BIP increases the cognitive load associated with distraction tasks.
    \end{itemize}
    \item \textbf{H3}: Presence and reaction time are correlated.
      \begin{itemize}[itemsep=-0.05cm, topsep=-0.1cm,leftmargin=0.35cm]
\item \textbf{H3a}: Presence mediates the relationship between cognitive load and reaction time.
    \end{itemize}
    
\end{itemize}

\subsection{Design}
For our study, we introduced distractions using secondary tasks~\cite{payne2024considerations-presence-secondary-tasks,bracken2014revisiting-reaction-time-secondary-task} rather than other common methods such as mimicking system glitches~\cite{breaking-experience}, whiteouts~\cite{slater2006analysis-physiological-BIP, garau2004temporal0slater-BIP-qualitative}, or interface interruptions~\cite{slater2000virtual-BIP-presence-counter}. Our decision stems from the need to create controlled and repeatable disruptions that allow us to evaluate their impact on our study variables. We assume that secondary tasks effectively simulate the types of interruptions users experience in real-world MR settings, though they may not capture spontaneous environmental changes.
Unlike system glitches or interface errors, which are often unpredictable and difficult to standardize across participants, secondary tasks provide a consistent and clear way to assess how participants reacted between the primary and secondary stimuli. This approach allows us to maintain experimental control by ensuring that each participant encounters distractions of the same nature and frequency, making comparisons across conditions more reliable.
Additionally, while glitches or sudden interruptions may cause frustration or cognitive overload, they could also introduce unintended variables such as frustration, confusion, or system usability issues that are difficult to disentangle from presence or BIP. In contrast, secondary tasks can be tailored to focus directly on cognitive load without introducing usability confounds~\cite{steed2005breaks-usablity-bip}. 

\subsubsection{Task Structure}
\noindent \textbf{\emph{Primary Task.} }
Participants complete an image sorting task across three conditions: No Distraction (\ND), \CD, and \ID. Each image is either a living or non-living object. The participant categorizes the image by dragging it to one of two labels: Living or Non-Living. This task remains consistent across all three conditions.
To ensure the images used do not trigger any unintended emotional responses (fear, disgust, confusion, etc.), we select 90 images from the Open Affective Standardized Image Set (OASIS)~\cite{kurdi2017-oasis-dataset}. This image set contains 900 images rated by 822 participants on pleasantness and excitement. We chose images with the highest ``pleasantness" ratings to minimize potential triggers.

\noindent \textbf{\emph{Secondary Task.}} \ND serves as the baseline; participants perform the primary image sorting task without distractions. In the \CD and \ID, participants perform secondary tasks. 
In the \CD, the primary task remains the same as in \ND but with an added congruent distraction. A virtual buzzer appears randomly between trails (details in \S\ref{sec:exp-task}), prompting the participant to turn off the buzzer by pressing a virtual button. The distraction is congruent because the button is integrated into VE, reducing the degree to which it disrupts the immersive experience. Therefore, while the task becomes more complex due to the secondary task, we hypothesize that the cognitive transition between sorting images and turning off the buzzer remains smooth because both actions occur in the same virtual space.

In the \ID, participants perform the primary task with a real-world distraction. A pop-up appears in the VE, but to turn off the buzzer, the participant presses a physical button in the real world outside the VE, located in the same physical position as the virtual button in the \CD. This forces the participant to break immersion to perform the real-world task. Unlike the \CD, \ID requires participants to repeatedly shift between the VE and real-world tasks, introducing a cognitive discontinuity that we hypothesize potentially increases cognitive load and reduces presence. We assume that requiring participants to interact with a real-world object physically reliably triggers a BIP event; however, individual variation in spatial awareness and task familiarity may modulate this effect

Moreover, as reaction time is a primary measure, we designed both tasks to be simple enough for any adult to perform, avoiding requiring specialized skills. This minimizes cognitive overhead, allowing us to focus on the effects of distractions without adding unnecessary complexity.

\begin{figure}
    \centering
    \includegraphics[width=0.4\linewidth]{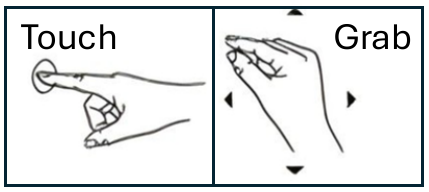}
    \vspace{-0.35cm}
    \caption{Hand gesture for virtual button press and dragging images.}
    \label{fig:hand-gesture}
\end{figure}
\subsubsection{Task Interaction}
\label{sec:task-interaction}
We use near interactions on the Hololens2~\cite{hand-manipulation-hololens2}, as several studies have confirmed the accuracy and reliability of its hand tracking~\cite{soares2021accuracy-hololens-hand-tracking, chandio-tvcg-23}, allowing participants to interact with virtual objects using natural hand gestures. 
For grab interactions (to move/drag the image in the primary task), participants use their thumb and index finger together. They perform a pinch gesture, pinching their thumb and index finger near a virtual object (image) to grab it, as shown in~\autoref{fig:hand-gesture}. Once the object is grabbed, participants can move their hand to move it to the desired location and release it by unpinching their fingers. For the virtual button press (distraction in \CD), we use touch interaction; participants touch their fingers near a virtual button, and the Hololens' hand tracking system detects this gesture, allowing the button to be pressed as it would in the real world, as shown in~\autoref{fig:hand-gesture}. The participants extend their index finger and bring it close to a virtual button. As the finger approaches, a slight pressing motion simulates a button press (poke gesture), and the system detects the interaction. 

To detect the real-world button press in the \ID, we used audio processing in our Unity app to capture the sound emitted by the physical button. When the button is pressed, it produces a distinct beep sound at a frequency of 220 Hz with a sine waveform. We processed the incoming audio using the Fast Fourier Transform (FFT) to analyze the frequency spectrum. The FFT algorithm identifies the 220 Hz beep, signaling that the button has been pressed. We also implemented a 10-second timeout for the buzzer, ensuring it automatically turns off if the signal is not detected or the button is not pressed in the VE (\CD) or the real world (\ID).

\begin{figure}[h]
    \centering
    \vspace{-0.3cm}
    \includegraphics[width=\columnwidth]{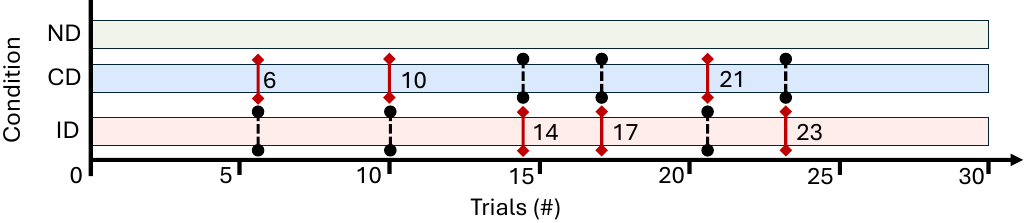}
    \vspace{-0.6cm}
    \caption{The two patterns of distractions for participants: red with diamond ends and black with circular ends. Each participant experiences (6, 10, 21) in one condition and (14, 17, 23) in the other.}
    \label{fig:distribution}
\end{figure}

\subsubsection{Order and Pattern of Distraction}
\label{sec:distraction-order}
To model the occurrence of distractions and mimic real-world unpredictability, we generated two distinct Poisson distributions. This maintains comparable difficulty and randomness across participants while preventing predictability or adaptation from using a single distribution. These distributions introduce distractions between the $6^{th}$ and $25^{th}$ trials, a boundary determined through the pilot study (\S\ref{sec:pilot}) to avoid early distractions. Using a Poisson process with a rate parameter of $\lambda = 0.15$, we generate exactly three distractions, subsampling any excess events to maintain randomness while controlling the number. Distribution 1 (trials $6, 10, 21$) is applied to the first distraction condition, and distribution 2 (trials $14, 17, 23$) to the other. This setup (shown in~\autoref{fig:distribution}) focuses on how the nature of distractions impacts our variables rather than their timing.

\section{User Study}
\label{sec:study}
\begin{figure*}[ht]
    \centering
    \includegraphics[width=\textwidth]{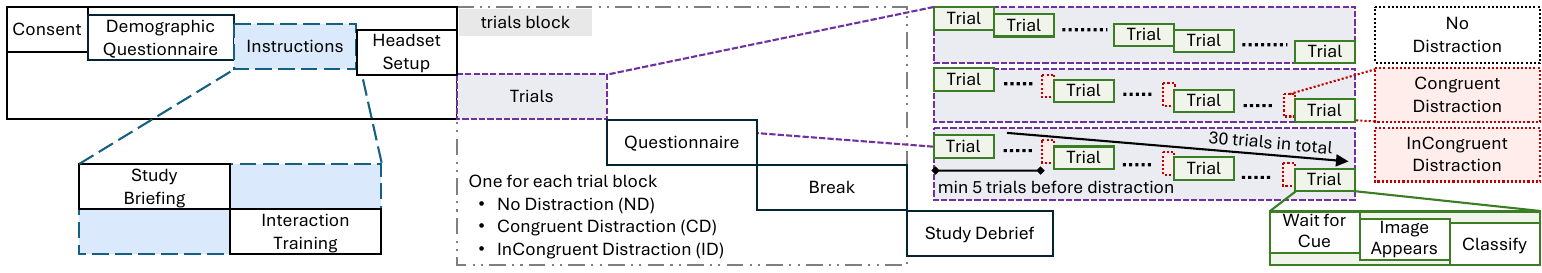}
    \vspace{-0.4cm}
    \caption{User study procedure.}
    \label{fig:study-procedure}
\end{figure*}
\subsection{Participants}
\label{sec:participants}
We conducted an a priori power analysis for our within-subject study comparing three conditions: \ND, \CD, and \ID with a medium effect size ($0.5$), an error probability $\alpha =0.05$, and a desired power of $0.95$ the analysis indicated a minimum of 43 participants. To account for potential attrition of 25\%, we recruited 54  participants from 18 to 35 years (M=23, STD=4.2), 26 were female, 1 non-binary, and 27 male. In terms of VR experience, 8 had never used it, 19 used it rarely, 20 occasionally, and 7 frequently. For the AR/MR experience, 21 had never used it, 16 rarely, 14 occasionally, and 3 frequently. Regarding gaming, 6 participants had no experience, 7 rarely played, 12 occasionally, and 29 frequently. 
All participants provided written informed consent, receiving \$15 for their participation. Each participant had normal or corrected-to-normal vision. The study was approved by the University of Massachusetts Amherst IRB, Human Research Protection Office.

\subsection{Material}
\label{sec:material}
We used the Hololens 2~\cite{hololens2} for its transparent display, spatial audio, integrated eye, spatial, and hand-tracking, and dual displays with a $1440\times936$ resolution and 110$\textdegree$ FoV with Qualcomm Snapdragon 850 CPU. The VE is implemented in Unity with XR plugins and custom scripts for hand tracking and recording reaction time.
Unity's Microphone API and audio processing are also used to detect the real-world button beep in the \ID to turn off the buzzer.

\subsection{Experimental Task} 
\label{sec:exp-task}
This section summarizes the cue, interaction, and feedback across all conditions in primary and secondary tasks.

\noindent \textbf{Cue.} In all conditions, during each trial of the primary task, an image to be sorted appears in the VE on a slow-moving trolley, prompting participants to sort it as living or non-living, as shown in ~\autoref{fig:teaser}. A virtual pop-up with a buzzer appears randomly in secondary task conditions (\ID, \CD). In \ID, the buzzer lacks spatial audio to mimic a real-world source, while in \CD, spatial audio signals that the sound originates from the VE.

\noindent \textbf{Interaction.} For the primary task, participants perform a pinch gesture to grab and drag the image to the appropriate label (living or non-living) in the VE. For the secondary task in \CD, participants press a virtual button with a poking gesture to stop the buzzer. In \ID, participants turn off the buzzer by pressing a physical button outside the VE on the physical table in front of them.

\noindent \textbf{Feedback.} In all conditions, once the image is sorted correctly, it disappears from the VE, signaling the completion of the primary task. For the secondary task, the buzzer stops when the virtual or physical button is pressed, resuming the primary image sorting task.

The equivalence between \CD and \ID lies in the interaction structure: both involve turning off a buzzer in the same location, requiring the same hand movement, with one in the VE and the other in the real world. We assume that participants interpret the physical and virtual button presses as functionally equivalent despite differences in sensory feedback, though additional factors such as tactile familiarity may influence their response.

\subsection{Measurement}
\label{sec:measuement}
We measure presence with the PQ and IPQ after each condition. The PQ evaluates factors such as the possibility to act (ACT) and examine (EXAM), realism (REAL), self-evaluation (EVAL), and interface quality (IFQUAL). The IPQ measures factors such as spatial (SP) and general presence (GP), realism, and involvement (INV). The presence scores are derived from 33 items (14 IPQ and 19 PQ) on a 7-point scale. To better fit the MR with minimal text change as validated by~\cite{plausible-tvcg-23,vona2024collabrating-AR-misalignment-element}, only one item was adapted, changing the “computer-generated world” to “augmented reality” (as it is more common term then MR, more details in \S\ref{sec:measuring-presence}). For each participant, presence was measured once at the end of each condition, resulting in 3 presence data points per participant (one for \CD, \ID, and \ND conditions). Higher scores indicate a stronger sense of presence, and lower scores suggest reduced presence.
Presence is used as a dependent variable to evaluate the effects of distraction type (\emph{\textbf{H1, H3}}), as well as an independent variable (\emph{\textbf{H1}}) and mediator in relation to cognitive load (\emph{\textbf{H2c}}) and reaction time (\emph{\textbf{H3a}}). 

In the absence of a widely validated BIP measurement tool, we used the same presence questionnaires (PQ and IPQ) as a proxy for BIP. Prior research supports using presence scores to infer BIP events, as significant drops in presence often indicate moments when the user’s attention shifts from the virtual to the real world~\cite{slater2000virtual-BIP-presence-counter}. This approach assumes that presence and BIP are inversely related; a decrease in presence correlates with an increased likelihood of BIP. While this method captures aggregate trends, it does not detect the precise moment a BIP occurs, which remains a limitation. 

Cognitive load is measured with the NASA Task Load Index (NASA-TLX)~\cite{hart1988development-NASA-TLX} after each condition. It is treated as a dependent variable (\emph{\textbf{H2, H2a, H2b, H2c}}) to assess the impact of distraction types and BIP on participants’ cognitive workload. Cognitive load also functions as an independent variable (\emph{\textbf{H3a}})  when exploring its influence on reaction time, with presence acting as a mediator.
Our software recorded reaction time on the HoloLens2 in milliseconds (ms) continuously during the image sorting task across all conditions (30 data points each). We analyzed reaction time as a dependent variable to examine its relationship with presence (\emph{\textbf{H3}}) and cognitive load (\emph{\textbf{H3a}}) in response to different types of distractions.

\subsection{Pilot Study}
\label{sec:pilot}
We conducted a pilot study with five participants to refine key parameters for the main experiment, including trial duration, frequency, timing and placement of distractions, and the selection of buzzer and beep (real-world) sounds and visual cues. We primarily relied on participant feedback to avoid experimenter bias, except for the trial duration, which was determined based on the reaction time data. The maximum reaction time recorded was 6.1 seconds, and the minimum was 2.2 seconds. To ensure consistency, we set the trial duration at 10 seconds, allowing ample time for all participants; however the extreme outliers may not fit this range. After testing trial counts between 20 and 40, we selected 30 trials to balance task engagement and minimize fatigue while providing sufficient opportunities for multiple distractions. We also tested various buzzer sounds and visual cues, choosing noticeable but not overly disruptive options. Based on feedback, the distraction timing was optimized to start from the $6^{th}$  trial and end at the $25^{th}$, with 3 (we tested 2-6) distractions being the most optimal among all the participants. We also fine-tuned the scale of images and labels for clarity. We adopted a slow-moving trolley mechanism for the image to appear in the scene, which participants found intuitive. Finally, we tested real-world button press beeps with two waveforms (sine, square) and frequencies ranging from 100 to 440 Hz to ensure reliable detection; participants preferred the 220 Hz beep with a sine wave.

\subsection{Study Procedure}
\label{sec:study-procedure}
The procedure for the study involved several steps, as shown in Figure~\ref{fig:study-procedure}.
 First, participants were given a consent form to read and sign before proceeding with the experiment. After consent was obtained, the participants were briefed on the experiments. During the briefing, participants were informed about the MR app and what to expect during the experiment. They were also instructed on the headset interactions and gestures needed to complete them, interactions required during the tasks, and how to perform them. The briefing included details about the visual stimuli used in the experiment, such as their color, shape, duration, cues, and feedback mechanism in each condition. Following the briefing, participants were asked to fill out a demographic questionnaire, which included questions about their gender, age, and familiarity with technology.
The core experiment consisted of three conditions: \ND, \CD, and \ID. These conditions were presented in a randomized and counterbalanced order using a Latin Square design to reduce any order effects across participants. In addition, a random distribution was assigned to the distraction tasks for each user, such that when Distribution 1 was used in the \CD, Distribution 2 was assigned to the \ID, and vice versa (\S\ref{sec:distraction-order}). This ensured that no participant experienced the same distribution for both distraction conditions, preventing adaptation and predictability in the task structure. Participants were then put on the headset, and each condition was presented one after the other, with a break to complete the questionnaires assessing their cognitive load and presence for each condition. 

In all three conditions, participants performed 30 trials (10 seconds apart) of the primary image sorting task, where they were shown a series of images and asked to categorize them as either ``living" or ``non-living" by performing a pinch-and-drag gesture. Participants completed two practice interactions before each condition, which were recorded but excluded from the analysis. In the \ND, participants performed the sorting task without any distractions, which took $\approx 5$ minutes to complete.
In the \CD, participants experienced a distraction in the form of a virtual pop-up accompanied by a buzzer. The pop-up appeared randomly during the trials, and participants were required to turn off the buzzer by performing a poking gesture to press a virtual button that appeared next to the pop-up, taking $\approx 6$ minutes to complete. In the \ID, participants faced a similar distraction, but this time, the buzzer sound came from the real world. Instead of pressing a virtual button, participants turned off the buzzer by pressing a physical button outside the VE on the table in front of them; in total, this condition took $\approx 6$ minutes. After completing all three conditions, participants were debriefed, and the purpose of the study was explained. The total duration of the session was $\approx 45$ minutes, including consent, briefing, training, headset calibration, the three experiments, three rounds of questionnaires, and debriefing.

\renewcommand{\arraystretch}{1.1}
\setlength{\tabcolsep}{4.2pt}
\begin{table}[t]
\scriptsize
\centering
\caption{Summary of statistics for presence scores (PQ + IPQ), NASA TLX assessment, and reaction time across conditions.}
\begin{tabular}{|l|c|c|c|c|c|c|c|c|}
\hline
& & &  & \textbf{Std.} & \multicolumn{2}{|c|}{\textbf{95\% CI (mean)}} & &  \\ \cline{6-7}
 & \textbf{N} & {\bf $\mu$} &  {\bf $\sigma$} & \textbf{Error} & \textbf{CI[5\%]} & \textbf{CI[95\%]} & \textbf{Min} & \textbf{Max} \\ \hline

\multicolumn{9}{|c|}{\textbf{Presence Score (PQ + IPQ, Scale: 0-7)}}\\ \hline
\textbf{\ND} & 54.0 & 4.65 & 0.85 & 0.12 & 3.18 & 5.75 & 2.59 & 5.83 \\ \hline
\textbf{\CD} & 54.0 & 4.34 & 0.97 & 0.13 & 2.59 & 5.73 & 2.21 & 6.18 \\ \hline
\textbf{\ID} & 54.0 & 4.19 & 0.88 & 0.12 & 2.56 & 5.35 & 2.29 & 6.52 \\ \hline \hline 

\multicolumn{9}{|c|}{\textbf{Cognitive Load (NASA TLX, Scale: 0-7)}}\\ \hline

\textbf{\ND}  & 54.0 & 1.62 & 1.13 & 0.15 & 1.0 & 4.67 & 1.0 & 4.80 \\ \hline
\textbf{\CD} & 54.0 & 2.92 & 1.13 & 0.15 & 1.33 & 4.80 & 1.00 & 5.00 \\ \hline
\textbf{\ID} & 54.0 & 4.03 & 1.12 & 0.15 & 1.26 & 4.80 & 1.00 & 6.00 \\ \hline \hline

\multicolumn{9}{|c|}{\textbf{Reaction Time (ms)}}\\ \hline
\textbf{\ND} & 54.0 & 3.90 & 1.45 & 0.20 & 1.45 & 6.22 & 1.0 & 6.67 \\ \hline
\textbf{\CD} & 54.0 & 5.50 & 2.21 & 0.30 & 2.26 & 8.93 & 1.0 & 8.52 \\ \hline
\textbf{\ID} & 54.0 & 7.88 & 2.54 & 0.35 & 3.15 & 10.57 & 1.0 & 9.72 \\ \hline\hline
\end{tabular}
\label{table:summary-stats}
\end{table}

\section{Results}
\label{sec:eval}
\subsection{Statistical Analysis}
\label{sec:result-stats-analysis}
Statistical analysis was performed using Python's Pingouin Library~\cite{Vallat-2018-Pingouin}. 
We first report detailed descriptive statistics for each condition (\ND, \CD, \ID) for presence scores (using combined PQ and IPQ scores), cognitive load (using NASA TLX assessment), and reaction time in~\autoref{table:summary-stats}. 
We used Q-Q plots to confirm the normality and Levene's test to verify the homoscedasticity for all conditions and variables. 
To examine the effect of condition on the presence score, we performed repeated measures ANOVA and reported the results in \autoref{table:anova-quest}. 
To further check the ANOVA results and understand the difference between conditions, we performed a post-hoc t-test and reported the statistics in \autoref{table:ttest-quest}.
Similarly, to understand the effect of condition on cognitive load, we performed repeated measures ANOVA and reported the results in \autoref{table:anova-tlx}. 
We also performed a post-hoc t-test for this case and reported the statistics in \autoref{table:ttest-tlx}.
We confirmed the sphericity of the data using the Mauchly test. 
Also, Cronbach's alpha value of 0.874 with a 95\%CI of [0.816, 0.917] demonstrated the high reliability and internal consistency of presence scores across conditions. 

\subsubsection{Presence Scores}
\label{sec:presence_scores_results}
The analysis of presence scores (a combination of PQ and IPQ) across the three conditions (\ND, \CD, \ID) reveals notable differences. Descriptive statistics (\autoref{table:summary-stats}) show the highest presence scores in the \ND condition ($M = 4.65, SD = 0.85$), followed by \CD ($M = 4.34, SD = 0.97$), and the lowest in \ID ($M = 4.19, SD = 0.88$). The 95\% confidence intervals for the means suggest differences across conditions, particularly between \ND and \ID, where there is no overlap between the intervals.
A repeated measures ANOVA (\autoref{table:anova-quest}) was conducted to assess these differences statistically. The results indicated a significant effect of condition on presence scores ($F(2, 106) = 12.51, p < 0.001, \eta^2 = 0.04$). This finding suggests that the type of distraction, whether congruent or incongruent, significantly affects a user's sense of presence in the MR.
To further explore these effects, post-hoc paired t-tests were conducted to compare presence scores between conditions (\autoref{table:ttest-quest}). The comparison between \ND and \CD showed a significant reduction in presence in the \CD condition ($t(53) = 3.14, p = 0.003, d = 0.34$), indicating that even \CD can negatively impact the sense of presence. The \ND vs. \ID comparison revealed an even more significant reduction in presence with \ID ($t(53) = 4.54, p < 0.001, d = 0.53$), demonstrating that incongruent distractions have a stronger detrimental effect on presence compared to congruent ones.
These results confirm that both types of distractions reduce presence, with \ID having a more pronounced effect. 

In \autoref{tab:subscales}, we show the values for each condition by summarizing the distribution of presence scores, where each participant contributes one score per condition. For both the PQ and IPQ subscales, participants in the \ND consistently reported higher presence scores compared to the \CD and \ID. 
In terms of the overall PQ score, participants in the \ND had the highest average presence score ($\mu = 5.46$, $\sigma = 0.89$), followed by \CD ($\mu = 4.88$, $\sigma = 1.02$) and \ID ($\mu = 4.21$, $\sigma = 0.90$). The same trend holds across the PQ subscales, where \ND consistently scores higher, particularly in the ACT subscale, with \ND having a mean of 5.71, while \CD and \ID drop to 4.69 and 4.24, respectively.
Similarly, the IPQ scores reflect a similar pattern. The \ND condition reports a mean presence score of 4.37 ($\sigma = 0.88$), with the \CD and \ID scoring 3.78 ($\sigma = 1.08$) and 3.34 ($\sigma = 0.85$), respectively. INV, IPQ-REAL, and SP also show higher scores for the \ND, which suggests that distractions substantially reduce participants' perceived presence.

\setlength{\tabcolsep}{5pt}
\renewcommand{\arraystretch}{1.08}
\begin{table}[t]
\centering
\scriptsize
\caption{Within subjects (\ND, \CD, \ID) repeated measures ANOVA using presence scores as the dependent variable. \textbf{SS:} Sum of Squares, \textbf{df:} Degrees of Freedom, \textbf{MS:} Mean Squares, \textbf{F:} F-values, \textbf{\emph{p}:} uncorrected $p$-value, \textbf{ng2:} generalized eta-square effect size, and \textbf{eps:} Greenhouse-Geisser epsilon factor.}
\vspace{-0.1cm}
\begin{tabular}{|l|c|c|c|c|c|c|c|}
\hline
\textbf{Source} & \textbf{SS} & \textbf{\emph{df}} & \textbf{MS} & \textbf{F} & \textbf{\emph{p}} & \textbf{ng2} & \textbf{eps}  \\ \hline

\textbf{Condition} & 5.88 & 2 & 2.94 & 12.51 & 0.00 & 0.04 & 0.93 \\ \hline
\textbf{Error} & 24.91 & 106 & 0.23 &  &  &  &  \\ \hline

\end{tabular}
\vspace{-0.3cm}
\label{table:anova-quest}
\end{table}

\begin{table}[ht]
\centering
\scriptsize
\caption{Paired samples t-test statistics for (\ND, \CD) and (\ND, \ID) using presence scores. \textbf{T:} T-value, \textbf{\emph{df}:} Degrees of Freedom, \emph{p}: $p$-value, \textbf{CI95\%:} confidence interval of the differences in means, \textbf{cohen-d:} Cohen d effect size, and \textbf{power:} Achieved power of the test.}
\begin{tabular}{|l|c|c|c|c|c|c|}
\hline
\textbf{Pair} & \textbf{T} & \textbf{\emph{df}}  & \textbf{\emph{p}} & \textbf{CI95\%} & \textbf{cohen-d} &  \textbf{power}  \\ \hline

\textbf{\ND, \CD} & 3.14 & 53 & 0.003 & [0.11, 0.50] & 0.34 &  0.68 \\ \hline
\textbf{\ND, \ID} & 4.54 & 53 & 0.000 & [0.26, 0.66]  & 0.53  & 0.97 \\ \hline

\end{tabular}
\vspace{-0.3cm}
\label{table:ttest-quest}
\end{table}

\subsubsection{Cognitive Load}
\label{sec:cognitive_load_results}
The cognitive load was measured using the NASA TLX assessment across three conditions (\ND, \CD, \ID). Descriptive statistics (\autoref{table:summary-stats}) show that cognitive load was lowest in the \ND ($M = 1.62, SD = 1.13$), increased in the \CD ($M = 2.92, SD = 1.13$), and was highest in the \ID ($M = 4.03, SD = 1.12$). The 95\% confidence intervals for the mean cognitive load across conditions indicate non-overlapping intervals, especially between \ND and \ID, suggesting significant differences between these conditions.
To statistically assess the impact of the distractions on cognitive load, a repeated measures ANOVA was performed (\autoref{table:anova-tlx}). The results revealed a significant main effect of condition on cognitive load ($F(2, 106) = 15.60$, $p < 0.001$, $\eta^2 = 0.08$), indicating that the type of distraction significantly influences cognitive load in the MR.
To further investigate these differences, post-hoc paired t-tests were conducted to compare cognitive load between conditions (\autoref{table:ttest-tlx}). The comparison between \ND and \CD showed a significant increase in cognitive load in the \CD ($t(53) = 3.94$, $p < 0.001$, $d = 0.60$), indicating that \CD imposes additional cognitive load compared to \ND. Similarly, the comparison between \ND and \ID  showed an even larger increase in cognitive load for \ID ($t(53) = 3.32$, $p < 0.001$, $d = 0.60$), highlighting the greater cognitive load caused by \ID.
These results demonstrate that both types of distractions significantly increase cognitive load, with \ID having a more pronounced impact. 

Although the accuracy of the task is not part of our hypotheses, ~\autoref{tab:accuracy} reports the accuracy of participants in sorting living and non-living images across three conditions. The highest accuracy is observed in the \ND (86.80\% mean accuracy), indicating that participants performed the sorting task most effectively when no distractions were present. In the \CD condition, accuracy drops to 74.92\%, suggesting that even \CD may reduce task performance, although participants still maintain relatively high accuracy. The lowest accuracy is seen in the \ID (52.42\%), where incongruence appears to have caused significant difficulty in performing the task. The increasing standard deviation from \ND (5.26\%) to \CD (14.85\%) and \ID (20.98\%) reflects greater variability in participant performance as distractions are introduced.

\begin{table}[t]
\scriptsize
\caption{Summary of descriptive results PQ and IPQ questionnaires subscales scores across conditions: No distraction (\ND), congruent distraction (\CD), and incongruent distraction (\ID). We provide mean ($\mu$) and standard deviation ($\sigma$) values for each experiment. The subscales for PQ and IPQ questionnaires are realism (PQ-REAL, IPQ-REAL), possibility to act (ACT), interface quality (IFQUAL), possibility to examine (EXAM), self-evaluation of performance (EVAL), involvement (INV), general presence (GP), and spatial presence (SP).}
\begin{center}
\begin{tabular}{||l|c|c|c|c|c|c||} \hline
{\bf Quest.} & \multicolumn{2}{|c|}{\textbf{\ND}} & \multicolumn{2}{|c|}{\textbf{\CD}} & \multicolumn{2}{|c|}{\textbf{\ID}}\\ \cline{2-7} 
& {\bf $\mu$} & {\bf $\sigma$} & {\bf $\mu$} & {\bf $\sigma$} & {\bf $\mu$} & {\bf $\sigma$} \\ \hline \hline \hline
PQ-REAL & 5.08 & 1.07 & 4.65 & 1.16 & 4.17 & 1.00 \\
\hline
ACT & 5.71 & 1.02 & 4.69 & 1.26 & 4.24 & 1.02 \\
\hline
QUAL & 5.62 & 1.06 & 5.03 & 1.01 & 4.30 & 1.34 \\
\hline
EXAM & 5.36 & 1.06 & 5.04 & 1.07 & 4.23 & 1.18 \\
\hline
EVAL & 5.85 & 1.16 & 5.58 & 1.45 & 4.12 & 1.54 \\
\hline
\textbf{PQ} & \textbf{5.46} & \textbf{0.89} & \textbf{4.88} & \textbf{1.02} & \textbf{4.21} & \textbf{0.90} \\
\hline\hline
INV & 3.90 & 1.12 & 3.26 & 1.22 & 2.91 & 0.98 \\
\hline
IPQ-REAL & 4.26 & 0.94 & 3.39 & 1.09 & 3.11 & 1.01 \\
\hline
SP & 4.72 & 1.13 & 4.27 & 1.42 & 3.78 & 1.19 \\
\hline
GP & 4.94 & 1.43 & 4.89 & 1.59 & 3.85 & 1.32 \\
\hline
\textbf{IPQ} & \textbf{4.37} & \textbf{0.88} & \textbf{3.78} & \textbf{1.08} & \textbf{3.34} & \textbf{0.85} \\
\hline \hline
\end{tabular}
\label{tab:subscales}
\end{center}
\vspace{-0.2cm}
\end{table}

\setlength{\tabcolsep}{5pt}
\renewcommand{\arraystretch}{1.08}
\begin{table}[ht]
\fontsize{6}{7}\selectfont 
\centering
\vspace{-0.1cm}
\caption{Within subjects (\ND, \CD, \ID) repeated measures ANOVA using cognitive load as the dependent variable. \textbf{SS:} Sum of Squares, \textbf{\emph{df}:} Degrees of Freedom, \textbf{MS:} Mean Squares, \textbf{F:} F-values, $p$: uncorrected $p$-value, \textbf{ng2:} generalized eta-square effect size, and \textbf{eps:} Greenhouse-Geisser epsilon factor.}
\begin{tabular}{|l|c|c|c|c|c|c|c|}
\hline
\textbf{Source} & \textbf{SS} & \textbf{\emph{df}} & \textbf{MS} & \textbf{F} & \textbf{$p$} & \textbf{ng2} & \textbf{eps}  \\ \hline
\textbf{Condition} & 16.72 & 2 & 8.36 & 15.60 & 0.00 & 0.08 & 1.00 \\ \hline
\textbf{Error} & 56.80 & 106 & 0.54 &  &  &  &   \\ \hline 
\end{tabular}
\vspace{-0.3cm}
\label{table:anova-tlx}
\end{table}

\begin{table}[ht]
\fontsize{6}{7}\selectfont 
\centering
\caption{Paired samples t-test statistics for (\ND, \CD) and (\ND, \ID) using NASA TLX. \textbf{T:} T-value, \textbf{\emph{df}:} Degrees of Freedom, \emph{p:} $p$-value, \textbf{CI95\%:} confidence interval of the differences in means, \textbf{cohen-d:} Cohen d effect size, and \textbf{power:} Achieved power of the test.}
\begin{tabular}{|l|c|c|c|c|c|c|}
\hline
\textbf{Pair} & \textbf{T} & \textbf{\emph{df}} & \textbf{\emph{p}} & \textbf{CI95\%} & \textbf{cohen-d} &  \textbf{power}  \\ \hline
\textbf{\ND, \CD} & 3.94 & 53 & 0.000 & [0.34, 1.03] & 0.60 &  0.99 \\ \hline
\textbf{\ND, \ID} & 3.32 & 53 & 0.000 & [0.27, 1.11]  & 0.60  & 0.99 \\ \hline
\end{tabular}
\label{table:ttest-tlx}
\end{table}
\subsubsection{Reaction Time}
\label{sec:reaction_time_results}
\begin{figure*}[t]
    \centering
    \begin{tabular}{cccc}
    \includegraphics[width=0.235\textwidth]{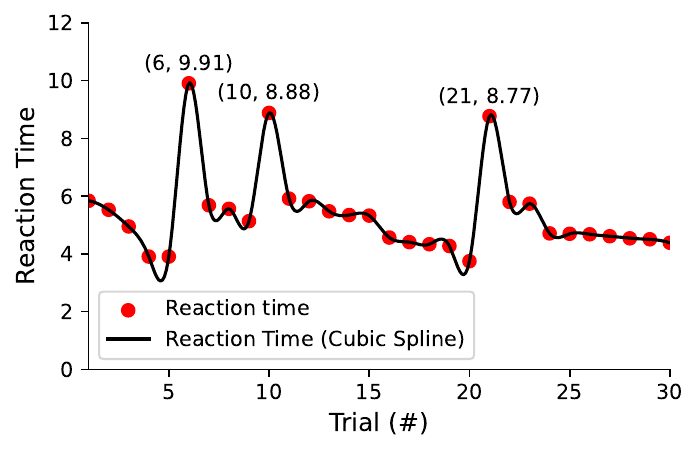} &
    \includegraphics[width=0.235\textwidth]{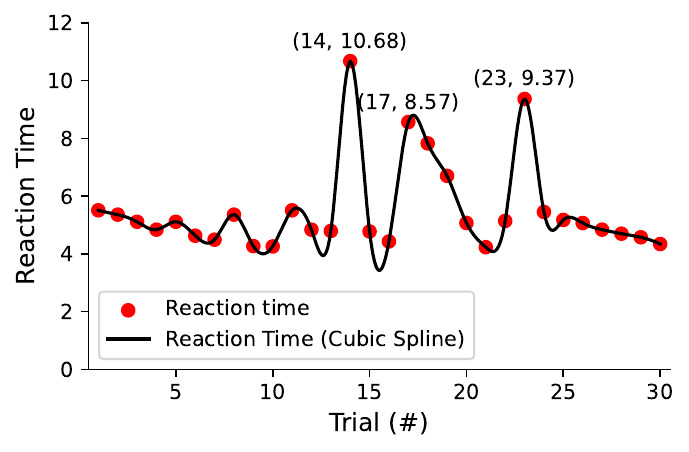} &
    \includegraphics[width=0.235\textwidth]{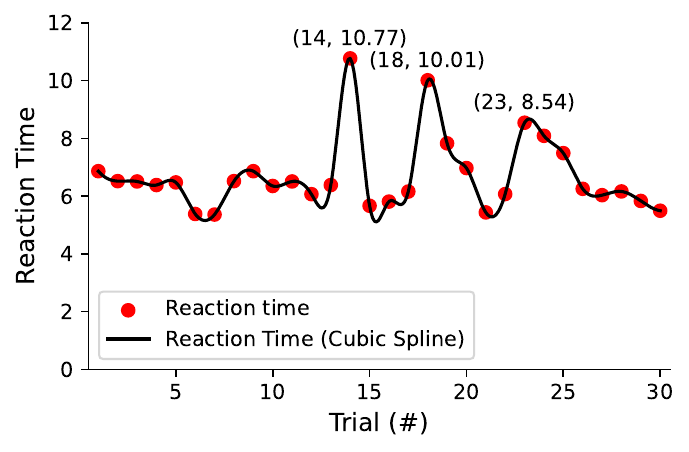} &
    \includegraphics[width=0.235\textwidth]{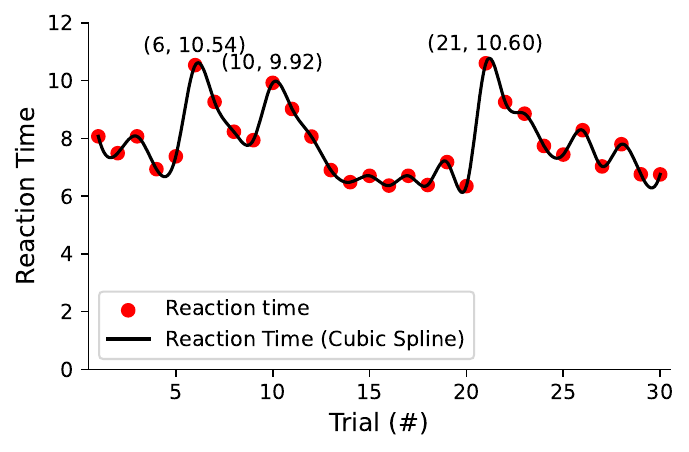} \vspace{-0.05cm} \\
    (a) \CD  (Distribution 1) & (b) \CD  (Distribution 2) & (c) \ID (Distribution 1) & (d) \ID (Distribution 2)  \\
    \end{tabular}
    \vspace{-0.4cm}
    \caption{\emph{Time series of reaction time across the two BIP conditions for participants with each distribution of distraction. In \CD, participants can recover easily from the low presence, whereas in \ID, they take much longer to recover from a break-in presence.}}
    \label{fig:reaction_time_timeseries}
\end{figure*}

The reaction time results indicate clear differences across all conditions. In the \ND, participants had the fastest average reaction time ($\mu = 3.90$, $\sigma = 1.45$), suggesting minimal interference from distractions. In the CD condition, the average reaction time increased significantly ($\mu = 5.50$, $\sigma = 2.21$), with some participants reaching up to 9.52 seconds, as depicted in the cubic spline model in~\autoref{fig:reaction_time_timeseries} (a,b). Finally, in the \ID, participants took the longest to react ($\mu = 7.88$, $\sigma = 2.54$), with recovery times remaining high across trials and individual times reaching 10.72 seconds in the distributions shown in~\autoref{fig:reaction_time_timeseries} (c,d). These results demonstrate that \ID substantially impacted reaction times, with recovery taking much longer compared to \CD and \ND. The longer reaction times here most likely reflect BIP-related cognitive load rather than external influences such as confusion or fatigue, although this distinction may not always be clear.

\begin{figure}[t]
    \centering
    \includegraphics[width=.72\columnwidth]{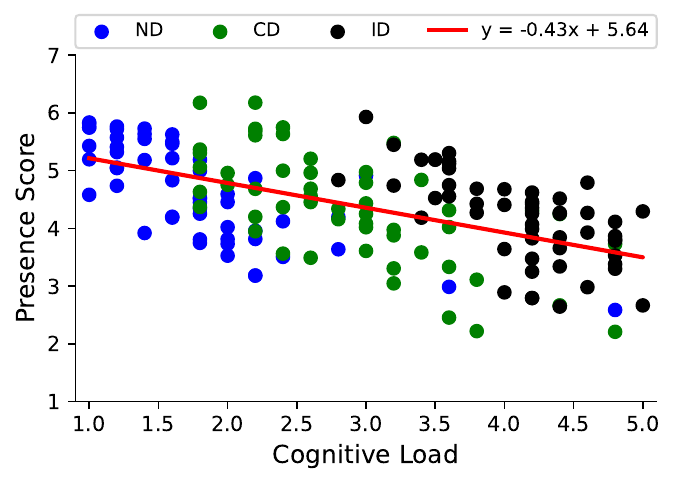}
    \vspace{-0.5cm}
    \caption{The relationship between cognitive load and presence. The regression coefficients are -0.43 and 5.64. The value of R$^2$ is 0.35 with a $p$-value of 0.000. The Pearson's correlation coefficients for \ND, \CD, \ID, and overall are -0.79, -0.73, -0.69, and -0.59, respectively. }
    \label{fig:tlx-quest-regression}
\end{figure}

\begin{figure}[t]
    \centering
    \includegraphics[width=.72\columnwidth]{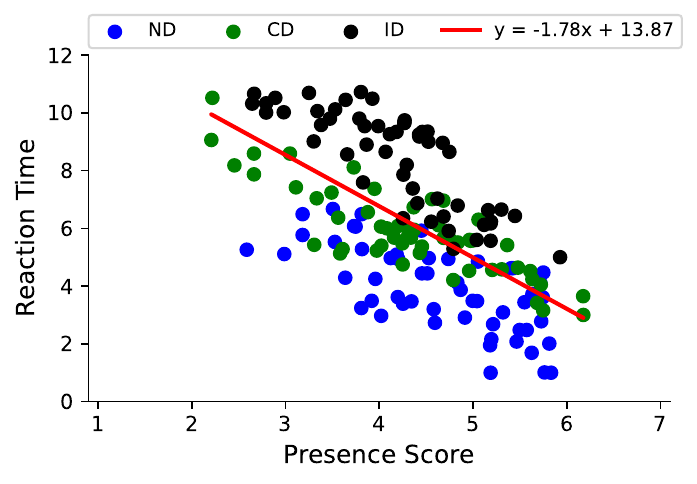}
    \caption{The relationship between presence score and reaction time. The regression coefficients are -1.78 and 13.87 with R$^2$ of 0.42 and $p$-value of 0.000. The Pearson's correlation coefficient is -0.68 for \ND, -0.84 for \CD, -0.80 for \ID, and -0.64 overall. }
    \label{fig:quest-rt-regression}
\end{figure}

\subsection{Correlation and Mediation Analysis}
\label{sec:correlation-mediation}
Next, we quantified the strength and direction of the relationship between cognitive load, presence, and reaction time.
First, we computed Pearson's correlation coefficient and applied linear regression to assess the relationship between cognitive load and presence across all conditions (\ND, \CD, \ID). As shown in \autoref{fig:tlx-quest-regression} a significant negative relationship exists between cognitive load and presence; 
with regression coefficients of $-0.43$ and $5.64$. 
The \(R^2 = 0.35\) and a $p$-value of $0.000$ indicate a strong negative relationship between cognitive load and presence. The Pearson’s correlation coefficients for the \ND, \CD, and \ID are $-0.79, -0.73, and -0.69$, respectively, with an overall correlation of $-0.59$. These results suggest that presence decreases across all conditions as cognitive load increases.
Similarly, we analyzed the relationship between presence scores and reaction time using Pearson's correlation and linear regression as shown in \autoref{fig:quest-rt-regression}, the regression coefficients were $-1.78$ and $13.87$, with an \(R^2 = 0.42\) and a $p$-value of $0.000$. This indicates a significant negative relationship between presence and reaction time, where higher presence scores are associated with faster reaction times. The Pearson's correlation coefficients for \ND, \CD, and \ID are $-0.68, -0.84, and -0.80$, respectively, with an overall correlation of -0.64.

\begin{figure}[t]
    \centering
    \includegraphics[width=.85\columnwidth]{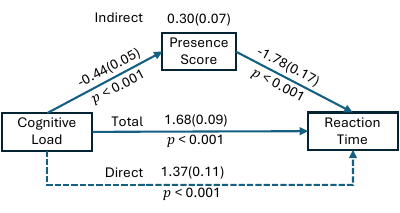}
    \vspace{-0.5cm}
    \caption{Bootstrap analysis (95\% CI 1000 resamples) for the mediation between cognitive load, presence, and reaction time, with a CI of [0.20, 0.42] for the indirect effect.}
    \label{fig:tlx-rt-mediation}
\end{figure}

We conducted a mediation analysis to understand how presence mediates the relationship between cognitive load and reaction time, shown in \autoref{fig:tlx-rt-mediation}. We computed both direct and indirect effects using this bootstrapping approach with 1000 resamples, and 95\% CI revealed an indirect effect of presence, with a CI of $[0.20, 0.42]$.
The total effect of cognitive load on reaction time was \( -1.78 \) with a standard error of \( 0.17 \) and a \( p < 0.001\), indicating a significant relationship between cognitive load and reaction time. As expected, this suggests that a higher cognitive load leads to slower reaction times.
The direct effect of cognitive load on reaction time, after accounting for the mediator (presence), was \( 1.68 \) with a standard error of \( 0.09 \) and a \( p < 0.001 \). This significant direct effect indicates that cognitive load still impacts reaction time independently of presence.
The mediation analysis revealed an indirect effect of \( 0.30 \), with a standard error of \( 0.07 \). The indirect pathway was significant (\( p < 0.001 \)), indicating that part of the effect of cognitive load on reaction time is mediated by presence. The coefficient of \( -0.44 \) with a standard error of \( 0.05 \) reflects that as cognitive load increases, presence decreases.

\begin{figure}[t]
    \centering
    \includegraphics[width=.8\columnwidth]{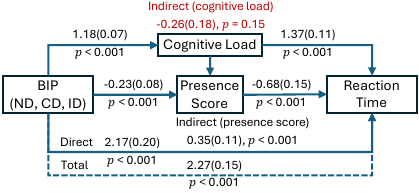}
    \caption{Overall mediation analysis with cognitive load and presence scores as the mediators for the relationship between different conditions and the reaction time.}
    \label{fig:mediation-overall}
\end{figure}

Lastly, the overall mediation analysis in~\autoref{fig:mediation-overall} shows that cognitive load and presence scores significantly mediate the relationship between BIP and reaction time. The total effect of BIP on reaction time is significant, as demonstrated by a coefficient of 
$2.17(0.20),p<0.001$, suggesting that different BIP conditions impact how long participants take to respond.

\begin{table}[ht]
\centering
\scriptsize
\caption{The accuracy of sorting living and non-living objects for study participants in \ND, \CD, and \ID conditions.}
\begin{tabular}{|l|c|c|c|}
\hline
\textbf{Stat} & \ND & \CD & \ID  \\ \hline
Mean ($\%$) & 86.80 & 74.92 & 52.42 \\ \hline
Std. Dev. ($\%$) & 5.26 & 14.85 & 20.98  \\ \hline
\end{tabular}
\label{tab:accuracy}
\vspace{-0.2cm}
\end{table}

\section{Discussion}
\label{sec:disc}
The study reveals how different distractions affect presence, cognitive load, and reaction time in MR. With medium effect sizes in cognitive load and presence score with high reliability (Cronbach's alpha = 0.874), we assume that our manipulation to create distractions influences presence and cognitive load to a certain extent. We conducted a thematic analysis to identify comments related to real-world distraction. 34 participants explicitly mentioned pressing the button on the table, indicating that our manipulations were noticed.

\subsection{Interpreting Results}
Our result partially support \emph{\textbf{H1}}. Both distraction types led to significant reductions in presence, but \ID had a more pronounced effect compared to \CD, as shown by the larger decrease in presence scores. Although we cannot directly measure BIP due to the lack of a validated tool, the reduction in presence scores suggests that BIP is more likely to occur in the presence of \ID, thereby partially supporting \emph{\textbf{H1}}.
For \emph{\textbf{H2}}, the findings fully support this hypothesis. The results show that both \CD and \ID significantly increased cognitive load compared to the \ND. The post-hoc t-tests indicate that \ID leads to a greater increase in cognitive load compared to \CD, supporting \emph{\textbf{H2a}}.
The evidence from the correlation analysis supports \emph{\textbf{H2b}}. The significant negative correlation between presence and cognitive load across all conditions, particularly in \ID, indicates that as presence decreases (a likely indication of BIP), cognitive load increases. Thus, \emph{\textbf{H2b}} is accepted based on these indirect indicators of BIP leading to increased cognitive load. The mediation analysis suggests that presence mediates the relationship between cognitive load and reaction time, supporting the indirect link between cognitive load and presence in \emph{\textbf{H2c}}. This finding shows that as cognitive load increases, it reduces presence, leading to slower reaction times.
\emph{\textbf{H3}} is supported by our data. There was a significant negative correlation between presence and reaction time across all conditions, indicating that as presence decreases, reaction time increases. \emph{\textbf{H3a}} is partially supported; while the correlation between presence and reaction time is strong, the mediation analysis shows that presence only partially mediates the relationship between cognitive load and reaction time. This means presence influences reaction time, but cognitive load also directly affects it beyond the mediation by presence; thus, we fully accept \emph{\textbf{H3}} and partially accept \emph{\textbf{H3a}}.

\subsection{Implications}
Our study makes several contributions to the field of MR research. First, we advance the understanding of how distractions impact core experiential factors such as presence, BIP, cognitive load, and reaction time. Our work extends existing theories of presence and cognitive load by showing how these constructs interact under varying distraction conditions, leading to different performance outcomes. This theoretical insight could inform the design of more immersive and user-friendly MR systems by identifying key variables that need to be managed to maintain a high presence and low cognitive load. We introduce a conceptual model (CDM-MR) that positions presence as a mediating factor between cognitive load and reaction time. The integration of presence, cognitive load, and reaction time in a cohesive framework represents a novel contribution, offering a theoretical foundation for future studies aimed at optimizing MR environments. Lastly, by providing empirical evidence of these relationships, we contribute to the growing body of literature on user experience in MR environments, highlighting the nuances of different types of distractions (congruent and incongruent).

\section{Limitations and Future Work}
\label{sec:future}
In this section, we briefly list our limitations, study scope elements that may appear as limitations and future directions for research. This includes balancing experimental control, the scope of the research for a single paper, focusing on specific hypotheses and addressing the practical constraints.

\subsection{Limitations}
First, we used secondary tasks as the primary method of inducing BIP events. While this approach effectively controlled distractions in a repeatable manner, it does not represent the full range of distractions present in real-world MR environments, such as spontaneous environmental noise or system glitches. We did not include these other types of distractions because we aimed to create a controlled experimental setup that allowed us to systematically measure the effects of task-related distractions without introducing too many variables that could affect the consistency of the results. 

Secondly, we used presence scores to infer BIP, which presents a limitation in accurately capturing BIP. This was necessary because no widely validated questionnaire or tool specifically designed to measure BIP exists. As a result, we had to rely on presence scores as a proxy, which may not fully capture the nuanced nature of BIP occurrences. However, for our study, the overall experience and the impact of BIP on cognitive load and reaction time were more critical than pinpointing the exact moments when BIP events occurred. Thus, while a more precise BIP measurement tool would have been ideal, the focus of our research on the broader effects of BIP justified the use of presence scores. 

Additionally, we employed a limited number of questionnaires to measure cognitive load and presence. This choice was made to keep the study manageable in scope and to minimize participant fatigue, as extensive questionnaires could have led to longer sessions and lower data quality due to participant disengagement. 
Lastly, our sample of younger adults from a college town may limit the generalizability of our findings due to its limited age diversity. This participant pool reflects the practical limitation of being in a college town, where diversity in age and experience is constrained. However, since age was not the focus of any hypotheses, we did not prioritize age diversity in the study.

\subsection{Study Scope}
We chose not to explore the potential reversible relationship between presence and reaction time. While reduced presence correlates with slower reaction times, investigating whether faster reaction times directly enhance presence was not relevant to our study's objectives. Since reaction time is already a well-established and objective metric, we focused on its direct relationship with presence rather than reversing the analysis. Exploring this reversibility would not have provided significant additional insights.
Another scope limitation is the inability to establish causality, as it requires more complex experimental designs such as longitudinal studies, randomized controlled trials, or interventions. These approaches were not feasible due to time, scope, and resource constraints. Additionally, the tools and methods needed to prove causality in the context of cognitive load, presence, and reaction time in MR environments are still evolving and may not be readily available or practical within the scope of this study.

\subsection{Future research}
In future research, we aim to address these limitations and expand the study's scope. Incorporating physiological measures along with the reaction time, like EEG, could provide real-time data on brain activity, offering more insights into cognitive states during BIP events and fluctuations in presence. This would allow researchers to understand better the neural correlates of attention shifts between virtual and real-world stimuli. In terms of distractions, future work should explore more diverse methods, such as environmental stimuli (e.g., fluctuating light levels, ambient noise), system-based glitches (e.g., lags, graphical errors), or sensory overload (e.g., haptic disturbances). Investigating these distractions will provide a more comprehensive understanding of how various real-world disruptions impact cognitive load, reaction time, and presence.

Future work should also explore the relationship between BIP and task-technology fit. Understanding how well MR systems align with users' tasks will offer valuable insights into presence, cognitive load, and reaction time. A better fit between task demands and technological capabilities could reduce cognitive strain and BIP events, thereby improving overall user performance and maintaining immersion. Studying this interaction will be essential for optimizing MR environments and ensuring that the technology effectively supports the user's goals.

\section{Conclusion}
\label{sec:conclusion}
This study examined the impact of distractions on reaction time, presence, and cognitive load in MR environments, focusing on BIP events. The results show that \ID increases cognitive load, slows reaction times, and reduces presence compared to \CD. These findings highlight the importance of studying distractions to better manage their effects on performance and presence. Reaction time is strongly correlated with presence, offering real-time insights into immersion, but further research is needed to develop advanced measurement methods to capture these complex constructs fully.

\acknowledgments{%
We thank the anonymous IEEE VR 2025 reviewers for their insightful comments and feedback. The research reported in this paper was sponsored by the National Science Foundation (NSF) under award 2237485.}
\bibliographystyle{abbrv-doi-hyperref}
\bibliography{main}

\end{document}